# Varying coupling constants and their interdependence

Rajendra P. Gupta
*Department of Physics, University of Ottawa*
*Ottawa, Canada K1N 6N5*
rgupta4@uOttawa.ca

**Abstract**

Since Dirac predicted in 1937 possible variation of gravitational constant and other coupling constants from his large number hypothesis, efforts continue to determine such variation without success. Such efforts focus on the variation of one constant while assuming all others pegged to their currently measured values. We show that the variations of the speed of light $c$, the gravitational constant $G$, the Planck constant $h$, and the Boltzmann constant $k$ are interrelated: $G \sim c^3 \sim h^{3/2} \sim k^{3/2}$. Thus, constraining any one of the constants leads to inadvertently constraining all the others. It may not be possible to determine the variation of a constant without concurrently considering the variation of others. We discuss several astrophysical observations that have been explained recently with the concomitant variation of two or more constants. We also analyze the reported and unexplained 35 $\mu$g decrease of 1 Kg Pt-Ir working standard over 22 years of measurement and show that the Kibble balance, that measures mass in units of Planck constant, cannot determine the variation of $h$ when $h$ and $c$ variations are interrelated as determined in here.

## 1. Introduction

Substantial theoretical and observational work has been done on determining the potential variation of the gravitational constant $G$ since Dirac in 1937 predicted its variation based on his large number hypothesis[1]. Teller in 1948 was the first to suggest a constraint on the variation of $G$ from the stellar scaling laws applied to the evolution of Solar luminosity and the environment required for the existence of life on Earth in the past[2]. Since then, many methods have been developed to determine the variation of $G$, which have all resulted in the constraints on $\dot{G}/G$ well below that predicted by Dirac. These include methods based on solar evolution[3], lunar occultation and eclipses[4], paleontological evidence[5], white dwarf cooling and pulsation[6], star cluster evolution[7], neutron star masses and ages[8], CMB anisotropies[9], big-bang nucleosynthesis abundances[10], asteroseismology[11], lunar laser ranging[12], evolution of planetary orbits[13], binary pulsars[14], supernovae type-1a (SNeIa) luminosity evolution[15], and gravitational wave observations of binary neutron stars[16].

While Einstein developed his ground-breaking theory of special relativity based on the constancy of the speed of light $c$, he did consider its possible variation[17] in 1907. Qi et al.[18] considered a power-law $c$ variation, and using the observational data of supernovae 1a (SNe1a), baryon acoustic oscillations (BAO), Hubble parameter $H(z)$, and cosmic microwave background, for very low and moderate redshift $z$ values, reported negligible $c$ variation. Another possibility of measuring the temporal variation of $c$ has been proposed by Salzano et al.[19]. They considered the relation between the maximum value of the angular diameter distance $D_A(z)$ and $H(z)$, and explored determining constraint on the variation of $c$ using the BAO and simulated data. Cai et al.[20] have used the independent determination of Suzuki et al.[21] of $H(z)$ and luminosity distance $D_L(z)$ from SNe1a observations to examine the variation of $c$. Cao et al.[22] reported the first measurement of $c$ value at $z = 1.7$ and found it to be essentially the same as measured on Earth, i.e., at $z = 0$. They used the angular diameter distance measurement for radio quasars extending to high redshifts. Cao et al.[23] have suggested a direct determination of $c$ variation using the galactic-scale measurements of strong gravitational lensing systems with SNe1a and quasars as the background sources. Lee[24] recently showed effectively no variation in the speed of light by the statistical analysis of a galaxy-scale strong gravitational lensing sample, including 161 systems with stellar velocity dispersion measurements.

Mendonca et al.[25] have tried to determine the variation of $c$ by using measurements of gas mass fraction of galaxy clusters with negative results.

Other constants of interest in our work are the Planck constant $h$ and the Boltzmann constant $k$. Mangano et al.[26] have studied the impact of time-dependent stochastic fluctuations of the Planck constant, and de Gosson[27] has applied the effect of a varying Planck constant on mixed quantum states. Dannenberg[28] has considered a possible temporal and spatial variation of the Planck constant by elevating it to the dynamical field that couples to other fields and to itself through the Lagrangian density derivative terms and studied the cosmological implications of such variations. He has also reviewed the literature on the subject. Direct measurement of the Boltzmann constant is based on the Doppler broadening of absorption lines in thermal equilibrium[29,30], such as the profile of rovibrational line of ammonia along a laser beam; the profile is related to the Maxwell-Boltzmann molecular velocity distribution through the kinetic energy contained in each molecule. Accordingly, it should be possible in principle to constrain the variation of the Boltzmann constant from a critical analysis of spectral line profiles of distant objects, such as quasars and interstellar media.

The statements above about the variability of constants have profound consequences when one tries to understand the true meaning of a constant, not in terms of the symmetry and Noether's theorem[31], but from a metrological perspective. How do we measure something, and in what units? Manmade units are arbitrary and have been defined historically for convenience using tools available. As measurement precision improved, the measuring tools, and hence the units, have evolved. As a result, it is natural to argue whether the variation of *dimensioned* constants, which obviously depends on the measuring tools, has any meaning. On the other hand, a *dimensionless* constant does not depend on the units used in its measurement. Thus any variation of a dimensionless constant truly represents new physics responsible for its variation. This subject has been of concern for time immemorial. It has been eloquently reviewed and discussed by Uzan[32,33] and Ellis and Uzan[34], among others. Duff[35,36] is an ardent advocate of dimensionless constants. He does not see any room for dimensioned constants, except as measuring tools that can define basic units of length, time, mass, etc., e.g., Planck units, Bohr units, Schrödinger units, Stoney units, Dirac units, and Natural units. Uzan[32,33] and Ellis and Uzan[34] have a more pragmatic approach. They are concerned about the consequence of the variation of constants on the equations of physics derived under the assumptions of their constancy: The constants must then be treated consistently as dynamical fields irrespective of whether they are dimensioned or dimensionless (Jordan 1937)[37]. Letting a constant vary in equations derived under the assumption of it being constant leads to incorrect results; one needs to go back to a Lagrangian that allows one to determine the degrees of freedom of the theory and check if it is well defined. They also have discussed the possible correlation among the variation of several dimensioned constants. Uzan (2011) explicitly states "…variations, if any, of various constants shall be correlated." in the context of the stability of the fine structure constant. Elsewhere in the same paper, he states, ".. it is necessary to relate the variations of different fundamental constants.".

Most experiments and observations study the potential variation of one constant while assuming all others pegged to their current value. This is inappropriate when several varying constants in the interpretation of data thus collected are involved. Eaves[38] has theoretically shown that the variation of the speed of light is related to the variation of the gravitational constant, and therefore constraining $G$ automatically constrains $c$. He has considered the distance measurement with the speed of light and came up essentially with the result that $G \sim c^3$, the same relation we have used in our earlier papers. Cuzinatto et al.[39] considered a scalar-tensor theory of gravity wherein the scalar field φ includes $G$ and $c$, both of which are allowed to be functions of the spacetime coordinates, and determined that $G \sim c^3$.

We have attempted to permit concurrent variation of $c, G, h,$ and $k$ in our cosmological, astrophysical and astrometric studies: $G \sim c^3 \sim h^3 \sim k^{3/2}$. As a result we were able to (i) resolve the primordial lithium problem[40], (ii) find an amicable solution to the faint young Sun problem[41], (iii) show that orbital timing studies do not constrain the variation of $G$[42], (iv) prove that gravitational lensing cannot determine the variation of $c$[43], and (v) establish that SNe Ia data and quasar data are consistent with the variable physical constants model[44,45].

The purpose of this paper is to explore the variation relation $G \sim c^3 \sim h^3 \sim k^{3/2}$. The local energy conservation relations are applied in Section 2 to derive the relation by studying core-collapse supernova explosion. We explore Kibble balance for testing the relation in Section 3, in Section 4 we discuss our findings, and report our conclusion in Section 5.

## 2. Interdependence of variation of the constants

Let us assume that the coupling constants evolve with the expansion of the Universe through scale factor $a$ as follows:

$$\text{Speed of light: } c = c_0 f_c(a).$$

Gravitational constant: $G = G_0 f_G(a)$.

Planck constant: $h = h_0 f_h(a)$.

Boltzmann constant: $k = k_0 f_k(a)$.

Here subscript 0 on a coupling constant refers to its current value, and the subscript on the arbitrary function $f(a)$ identifies the associated coupling constant.

Consider now an exploding star of mass $M$ and radius $r$, such as a core-collapse supernova, where a fraction $\eta$ of the mass is converted through fusion ($\eta \sim 0.7\%$ for hydrogen to helium conversion) into the nuclear energy causing the explosion. Assume a fraction $\beta$ of the explosion energy is used up in countering the negative self-gravitational energy of the mass to bring it to zero and the balance shows up as kinetic energy of the exploded particles (ignoring energy loss due to escaping neutrino and antineutrino particles). A fraction $\gamma$ of this kinetic energy thermalizes and is partially radiated away as photons. When distances are measured using the speed of light[38,42], the evolution of the energies may be written

$$\eta M c^2 \times \beta = \frac{GM^2}{r} = \frac{GM^2}{r_c(c/c_0)} \Rightarrow \eta \beta c^3 = \frac{GMc_0}{r_c}, \quad (1)$$

where $r_c$ is the stellar radius independent of the speed of light (similar to the comoving distance in cosmology) defined by $r \equiv r_c(c/c_0)$. Thus,

$$\eta \beta c_0^3 f_c(a)^3 = \frac{G_0 f_G(a) M c_0}{r_c}. \quad (2)$$

Local energy conservation over each slice of the cosmic time, i.e., scale factor $a$, leads to $f_G(a) = f_c(a)^3$, i.e., $G \sim c^3$.

Now consider the thermalized kinetic energy of $N$ particles, comprising mass $M$, at temperature $T$. Then

$$\eta(1-\beta)Mc^2 \times \gamma = NkT. \quad (3)$$

This means that $c^2 \sim kT$. Since $T$ is an arbitrary measure of thermal energy, $k \sim c^2$, i.e., $f_k(a) = f_c(a)^2$.

Finally, consider that a fraction $\delta$ of the thermal energy generates $N_\nu$ number of photons of frequency $\nu$. Then

$$\delta N k T = N_\lambda h \nu. \quad (4)$$

Since $N$ and $N_\nu$ are conserved in an evolutionary (expanding) universe, and $\nu$ may be measured in arbitrary unit of time, we must have $k \sim h$. But $k \sim c^2$, which leads to $h \sim c^2$, i.e., $f_h(a) = f_c(a)^2$.

In summary,

$$f_c(a) = f_h(a)^{1/2} = f_k(a)^{1/2} = f_G(a)^{1/3} \equiv f(a), \text{ or}$$

$$c = c_0 f(a), G(a) = G_0 f(a)^3, h(a) = h_0 f(a)^2,$$
$$\text{and } k = k_0 f(a)^2. \quad (5)$$

If $\eta, \beta, \gamma$, and $\delta$ are functions of the scale factor $a$, then they can be absorbed in functions representing the variations of the constants without affecting our findings.

The above has general applicability and is not confined to the supernovae explosions. The supernova explosion was chosen as it involves all the four types of energy conversions needed for the analysis. Also, it is easy to see that including the energy loss from escaping neutrinos and antineutrinos does not affect our study.

As pointed out by John Hunter (personal communication) the above can be represented in terms of the variation of the length dimension of each constant. Thus, every quantity varies according to its length dimension, i.e., if a quantity $Q$ has a length dimension of $n$, it's $f_Q(a) = f(a)^n$. It greatly simplifies the application of the variation of coupling constants to practical problems.

## 3. Experimental testing of the interrelationship

The coupling constants values have been historically determined from observation and experiments, not from any theory. Therefore, their variation must also be measurable and determined similarly. Singh et al.[59] have summarized the historical measurement data on several constants, including $c$, $G$, and $h$. The variation shown in such data is primarily due to the improvement in the precision of acquiring data. However, they feel there may be a statistical trend in the data variation. High precision data acquired over a period of ten to twenty years may reveal true trends in the variation or lack thereof within the limits of experimental precisions. Since most of the time, methods and equations used for measuring constants involve other constants, one has to be mindful of the possible covariation of multiple constants.

While we have recently used the relationship $G \sim c^3 \sim h^3 \sim k^{3/2}$ to analyse and fit astrophysical observations[40-46] instead of $G \sim c^3 \sim h^3 \sim k^{3/2}$ determined here, there are many other models which fit the same observations. Since the relationship determined here for Planck constant is different, the earlier findings will need to be reevaluated.) Thus, such fits are essential but not sufficient to unequivocally prove the relationship. We need to look for some high precision terrestrial or astrometric measurements that could provide a decisive test. After analysing various possibilities, we considered very high precision measurements of the Planck constant using the Kibble balance.

A Kibble balance[47,48] – also called watt balance and Planck balance – is an extremely high precision metrological weighing machine with an accuracy of up to a few parts per billion. In it the weight of a test mass is exactly compensated by a force produced by an electric current running in a coil surrounding the mass and a high field strength magnet surrounding the coil. The current in the coil is measured with a resistor in the circuit with its resistance determined to an accuracy of one part in a billion using quantum Hall effect, and the voltage driving the current is measured with an uncertainty of one part in ten billion using Josephson effect. The coupling constants that enter in the equations determining the mass are[49] $c$, $G$, and $h$. Assuming $c$ and $G$ are precisely known, the Kibble balance realizes the definition of the kilogram unit in terms of the Planck constant $h$. Some of the most accurate Kibble balances are NIST-4 at the National Institute of Standards and Technology, U.S.A.[49,50] and at the National Research Council of Canada[51,52]. Kibble balances are able to determine $h$ with a precision of few parts per billion ($h$ =6.626 070 15(12) × $10^{-34}$ J s).

Following Haddad et al.[49], the Kibble balance test-mass $m$ is proportional to $h/c^2$ [60]. Thus, using equation (5) $m \propto f(a)^0$, resulting in $\dot{m}/m = 0$. Now there was an unexplained decrease of an average 35 $\mu$g in ten 1 Kg Pt-Ir working standards[50,53,54] over a period of 22 years (1992 to 2014). It was a result of a calibration campaign of the working standards that are used routinely for the dissemination of the unit of mass by the BIMP (Bureau International des Poids et Mesures, France) against IPK (International Prototype of the Kilogram). If the mass decrease is confirmed by Kibble balance, it would translates to $\dot{m}/m = -1.6 \times 10^{-9}$ yr$^{-1}$. However, it cannot be accounted for due to the interrelated variation of coupling constants studied here.

There may be other laboratory experiments that could provide further test of the constants' interrelationship, such as the XRCD (x-ray crystal density) method for determining the Planck constant with extreme precision[55] similar to the Kibble balance. One purpose of this paper is to invite ideas for such experiments.

## 4. Discussion

We see from Section 2 that if we do not permit one of the coupling constants to vary, the others are naturally constrained not to vary. For example, if the Planck constant's variation is ignored, i.e., if $f_h(a)$ is set equal to 1, then $f(a)$ functions for all other coupling constants are automatically constrained to 1. This finding is very general and independent of any model or form of the function $f(a)$, including $f(a)$ to be a constant. It is the major problem in all the studies trying to determine constraints on the variation of one of the coupling constants, mostly $G$ or $c$, while considering all other constants to remain fixed at their current values. Thus, any limits determined on the variation of the selected constants in all the studies, such as those cited in Sec. 1, may be due to the limitations and errors in the observations and experiments involved.

Interestingly, the relationship among the variation of the constants was established in a somewhat circuitous way in earlier works[46]. The scaling $G \sim c^3$ was derived from the null results on the variation of the gravitational constant[12], $h \sim c$ from tight constraints on the variation of the fine structure constant[56], and $kT \sim ch$ from the blackbody radiation analysis. Based on the findings of this paper, some of these relationships will need to be corrected: $h \sim c^2$ and $k \sim h$ while still be able to retain the constancy of fine structure constant[60]. It is worth reiterating that very recently Eaves[38] and Cuzinatto et al.[39] showed $G \sim c^3$.

The approach we have taken is rather phenomenological. A more appropriate and satisfying approach would involve writing an action for the interrelated constants represented as fields and derive the relationship between $G \sim c^3 \sim h^{3/2} \sim k^{3/2}$ with the scale factor $a$. As suggested by Dannenberg[57], for $G, c,$ and $h$, one should experiment with couplings between Klein-Gordon action and scalar fields representing the three involved constants. It is not an easy task and we have initiated collaboration for this challenging work.

Regarding dimensioned and dimensionless constants, the constant $Gm_e^2/hc$ ($m_e$ being the electron mass) is dimensionless and can be considered resulting from the atomic clocks varying relative to gravitational clocks[34,58]. It should be constant if the two times do not evolve differentially. The findings of this paper corroborate this constancy.

We have discussed in Section 1 how the concurrent variation of several constants is able to explain several astrophysical observations[40-46]. Since other models can also explain the same observations, a laboratory experiment is required to test the relationship among the constants determined in this work. We have shown in Sec. 3 that the Kibble balance that measures the Planck constant with a precision of a few parts per billion does not have the potential of verifying or falsifying our findings.

## 5. Conclusion

We should not expect the variation of a constant when any of the remaining four constants considered in this work are pegged to their current value. Several astrophysical observations have been explained in our earlier studies using the interrelationship of the constants, indirectly validating it, but they can also be explained with alternative models. Thus, laboratory experiments need to be designed to test the relationship. Existing high precision measurement and mass calibration show an unexplained 35 $\mu$g average loss of the 1 Kg Pt-Ir working standards over 22 years. We expected this tiny mass loss could possibly be accounted for using the interrelationship among the constants. We have shown that this is not possible since variation of $h$ is cancelled by the variation of $c^2$. If one considered the variation of $h$ and $c$ independently then one would expect to be able to determine $h$ variation by Kibble balance.

## Acknowledgements

The author is grateful to Dr. Rodrigo Cuzinatto and Dr. Barry Wood for discussions, Dr. John Hunter, Dr. Rand Dannenberg, Dr. Reuben Eaves, and Dr. Yousef Bisabr for critical comments, and to Macronix Research Corporation for financial support. Special thanks are due to Dr. Stephan Schlamminger for considering the potential of the Kibble balance for checking the constants' variation and for informing decrease in the masses of 1 Kg Pt-Ir working standards over years, and to the reviewers for suggesting important modifications.